\documentclass{article}
\usepackage[intlimits]{amsmath}
\usepackage{amsfonts,amscd,cite}
\usepackage{bbm}

\usepackage{mathrsfs}

\DeclareSymbolFont{EulerExtension}{U}{euex}{m}{n}
\DeclareMathSymbol\intop  \mathop {EulerExtension}{"52}
\DeclareMathSymbol\ointop \mathop {EulerExtension}{"48}

\let\rt=\right
\let\lf=\left

\let\dl=\delta

\let\bt=\beta
\let\th=\theta
\let\gm=\gamma

\let\to=\rightarrow
\let\p=\partial 
\let\la=\langle
\let\ra=\rangle

\newcommand{\zz}{{\boldsymbol{z}}}

\newcommand{\eq}[1]{eq.~(\ref{#1})}
\newcommand{\ket}[1]{{|}\,#1\,{\rangle}}
\newcommand{\bra}[1]{\langle\,#1\,|}
\newcommand{\braket}[2]{\langle\,#1\,|\,#2\,\rangle}
\newcommand{\no}[1]{:\!#1\!:}
\newcommand{\pco}{\Gamma}
\newcommand{\ipco}{Y}
\newcommand{\hpco}{\Bar{\pco}}
\newcommand{\ihpco}{\Bar{\ipco}}

\newcommand{\ber}{\mathop{\rm Ber}\nolimits}
\newcommand{\rank}{\mathop{\rm rank}\nolimits}
\newcommand{\D}{{\mathcal{D}}}

\newcommand{\IC}{{\mathbb{C}}}
\newcommand{\IR}{{\mathbb{R}}}

\newcommand{\Lie}{{\mathcal{L}}}
\newcommand{\inn}{{\mathbf{i}}}
\newcommand{\ext}{{\mathbf{e}}}
\newcommand{\dd}{{\mathbf{d}}}
\renewcommand{\d}{{\mathrm{d}}}

\newcommand{\vv}{\boldsymbol{v}}
\newcommand{\xx}{{x}}
\newcommand{\st}{{t}}

\newcommand{\ov}{\hat{v}}
\newcommand{\cv}{{v}^\vee}
\newcommand{\ocv}{\hat{v}^\vee}
\newcommand{\cvf}{\Tilde{v}^\vee}

\newcommand{\cvv}{\boldsymbol{v}^\vee}

\renewcommand{\P}{{\mathscr P}}

\title{{}\hfill
    \raisebox{2cm}[0pt]{
      \begin{array}[b]{l}
        \mbox{\small IFP/104/UNC}\\
        \mbox{\small\tt hep-th/9706033}
      \end{array}
      }\\
    Picture changing operators in supergeometry and superstring
  theory}
\author{Alexander Belopolsky\thanks{\tiny Supported in part by the US
      Department of Energy under Grant DE-FG 05-35ER40219/Task~A.}\\
    University of North Carolina\\
    at Chapel Hill\\
    E-mail: \texttt{belopols@physics.unc.edu}}
\begin{document}
\maketitle
\begin{abstract}
  Geometrical meaning of superstring pictures is discussed in details.
  An off-shell generalization of the picture changing operation and
  its inverse are constructed.  It is demonstrated that the
  generalised operations are inverse to each other on-shell while
  off-shell their product is a projection operator.
\end{abstract}
\section{Introduction and summary}
\setcounter{equation}{0}
\label{sec:intro}

The notion of pictures was introduced in the early years of
superstring theory.  Originally pictures appeared as different Fock
spaces (\(F_1\) and \(F_2\)) which can be used to represent string
states \cite{nst71,bors73}.  The spectrum of physical states was shown
to be identical in both pictures and the rules for calculating
physical amplitudes required to take some states from \(F_1\) and
others from \(F_2\).  It was more than a decade later when the
physical origin of the pictures was clarified.  In their fundamental
paper \cite{fms86} Friedan, Martinec and Shenker (FMS) demonstrated
that the choice of a picture is equivalent to a choice of ghost
vacuum.  They showed that inequivalent representations of bosonic
superconformal ghosts can be built using vacuum states with different
Bose sea level.  Each of these ghost representations would lead to a
different BRST complex and therefore to a different description for
the physical states.  The picture changing operation was introduced as
a regular procedure for mapping the physical states (BRST cohomology)
from one description to another.

The FMS approach was based on the bosonized description of
superconformal ghosts and it provided a very efficient algorithm for
calculating the string amplitudes.  Although the bosonization formulae
can greatly simplify the calculations, they tend to obscure the
physical meaning of the operators. Thus in addition to bosonic ghost
fields \(\bt\) and \(\gm\), the FMS formalism introduces auxiliary free
fermions \(\eta\) and \(\xi\) and fermionic ``solitons'' \(e^{\phi}\)
and \(e^{-\phi}\) whose origin was rather unclear.  It was later
realized that the solitons can be interpreted as delta functions of
the bosonic ghosts \(e^{\phi}=\dl(\bt)\) and \(e^{-\phi}=\dl(\gm)\)
and they appear in the expressions for the string amplitudes due to
the gauge fixing in the functional integral \cite{ver87,DPol95}.  One
of the auxiliary fermions, \(\xi\) has also received a simple
interpretation as a Heavyside step function of the bosonic antighost
field \(\xi=\Theta(\bt)\).  This elegant interpretation explained the
operator product expansions for these fields, but it was still not
clear why a delta or step function of a bosonic field produced a
fermion \cite{am90}.

The existence of the inverse picture changing operation was
conjectured in the original paper of FMS and the corresponding
operator was constructed explicitly by Witten in order to formulate
the string field theory of open superstrings \cite{Witten86}.  Witten
used the operator \(\ipco(z)=\no{c(z)\p\xi(z)e^{-2\phi(z)}}\) as an
inverse to the FMS picture changing operator \(\pco(z)=[Q,\xi(z)]\).
Although a formal proof was presented in the original paper, the
existence of the inverse to the picture changing operation remained
controversial since the picture changing operation was known to change
the matter contents of a state while \(\ipco(z)\) was expressed in
terms of ghost fields exclusively. Recently there was an intriguing
attempt to construct a physical state which exists in one picture but
not in the other \cite{DPol97,DPol97a}.  An existence of such state
would definitely be an obstruction to the inverse picture changing
operation.  A closer look at the proposed state reveals that it is not
annihilated by the full superstring BRST operator\footnote{I would
  like to thank Walter Troost for sharing with me this observation.},
but it might be possible to modify the theory so that it is still
a physical state at least for the zero momentum\cite{DPol-pri}.

The picture changing operation was an essential ingredient of the FMS
formalism.  At least two different pictures were necessary in order to
calculate string amplitudes and to define the SUSY generators.  It was
long suspected that the need for different picture was an artifact of
a particular choice of odd coordinates on the super moduli space
\cite{ag88-2,Gid92}, but until recently there was no formalism
available to express the string amplitudes without first making a
choice of odd coordinates.  Such formalism was developed by the author
in \cite{bel96a,bel97} using the geometrical theory of integration on
supermanifolds.  In the latter paper an analog of the FMS picture
changing operation was introduced on the de Rham complex of a
supermanifold.  Since the FMS picture changing operation proved to be
unnecessary in the new formalism, it was not considered in
\cite{bel97} and the relation between the picture changing operations
on de~Rham and BRST complexes was not properly addressed.  In this
paper we will attempt to close this gap.

The paper is divided into three parts.  The first part is devoted to
picture changing operators in supergeometry, the second to picture
changing operators in superstring theory and the third provides a
bridge between the two applications.

The most important lesson which the supergeometry provides for the
string theory is that there are two ghost numbers, an even one and an
odd one, in the ghost sector.  The ghost number used in the FMS
approach is the difference of these two numbers and does not have a
separate geometrical meaning.  Both fermionic and bosonic ghosts
change only the even ghost number and one has to introduce the
solitons \(\dl(\bt)\) and \(\dl(\gm)\) in order to change the odd
ghost number.

The odd ghost number is closely related to the notion of pictures.
For each odd ghost number we can construct a separate BRST complex
closed under the action of the ghost algebra and the BRST operator. We
will show that although these complexes are not isomorphic, an
isomorphism exists in cohomology and can be provided by picture
changing operators.  Picture changing operators and their inverses can
also be extended off-shell, but their product will no longer be the
identity but a projection operator.  It is possible that a projection
to the subspace where the picture changing operation is invertible is
necessary for a successful construction of a superstring field theory.

\section{Picture changing operators in supergeometry}
\setcounter{equation}{0}
\label{sec:geom}

In this section we introduce a variant of a super generalization of
the de Rham complex of differential forms.  For a supermanifold of
dimension \(n|m\) we will define what will be called singular
\(r|s\)-forms.  Possible values for the odd degree of a singular form
are \(s=0,\dots,m\) while the odd degree, \(r\), can be any integer
number if \(s\) is not equal to \(r\) or \(m\); \(r\geq0\) for \(s=0\)
and \(r\leq n\) for \(s=m\).  The super de Rham complex,
\(\Omega^{\cdot|\cdot}=\bigoplus_{r,s}\Omega^{r|s}\) will come
equipped with a differential \(\dd:\;\Omega^{r|s}\to\Omega^{(r+1)|s}\)
and the cohomology will be defined for each permissible value of the
degree \(r\) and \(s\).  Since the differential does not change the
odd degree of a form, the de Rham complex naturally splits into
\(m+1\) independent subcomplexes each with a different value of \(s\).
We will refer to these subcomplexes as pictures.  Given a
non-vanishing odd vector field \(\hat{v}\) we will construct an
operator \(\pco_{\hat{v}}\) which will map forms in picture \(s\) to
forms in picture \(s-1\) and commute with the differential.
Composition properties of the picture changing operators will be
addressed. The existence of the picture changing operators strongly
suggests that the cohomology is the same in every picture.  We will
demonstrate this for a simple case of a \(1|1\) dimensional
supermanifold for which we will construct an inverse to the picture
changing operator on the cohomology.

\subsection{Singular differential forms on supermanifolds}
\label{sec:dfs}

Differential \(r|s\)-forms on the supermanifolds were first introduced
by Voronov and Zorich in \cite{vz86,vz86a,vz87,vz88,ThVor91}.  One of
the difficulties in their approach was that explicit examples were
difficult to obtain. Here we will use their definition with a slight
modification to relax the non-singularity condition.  By doing that we
will be able to produce plenty of forms which were not allowed in the
original works.  One of the major benefits will be an explicit
description of a large subset of differential forms on supermanifolds
closed under the action of the de Rham differential.   

A function of \(r\) even and \(s\) odd tangent vector to
a supermanifold,
\begin{displaymath}
  \omega^{(r|s)}(\vv)=\omega^{(r|s)}(v_1,v_2,\dots,v_r|
  \hat{v}_1,\hat{v}_2,\dots,\hat{v}_s),
\end{displaymath}
is called a differential \(r|s\) form if it satisfies the following
two conditions:
\begin{equation}
  \label{ber}
    \omega^{(r|s)}(J\,\vv) = \ber J\,\omega^{(r|s)}(\vv),\quad
 \mbox{for any \(J\in GL(r|s)\)}.
\end{equation}
and 
\begin{equation}
  \label{sform}
  \frac{\p^2\omega^{(r|s)}(\vv)}{\p v_F^A\p v_G^B}+(-1)^{[F][G]+[B]([F]+[G])}
  \frac{\p^2\omega^{(r|s)}(\vv)}{\p v_G^A\p v_F^B}=0,
\end{equation}
where we use the symbol \(\vv=(v_1,v_2,\dots,v_r|\hat{v}_1,\hat{v}_2,
\dots,\hat{v}_s)\) to refer collectively to the arguments of
\(\omega^{(r|s)}\). Naturally \(v^A_F\) denotes the \(A\)-th
component of the \(F\)-th vector, so indices \(F\) and \(G\) run from
\(1\) to \(r|s\) and \(A\) and \(B\)---from \(1\) to the dimension of
the manifold.  We will denote the space of differential \(r|s\)-forms
on the supermanifold \(M\) by \(\Omega^{r|s}(M)\). 

The first identity, \eq{ber} is called the invariance condition and
due to this condition we can define an integral of \(\omega^{(r|s)}\)
over a dimension \((r|s)\) submanifold \(N\in M\) 
\begin{displaymath}
  \int_N\omega^{(r|s)}=
\int_N\D(t^1,\dots,t^r|\hat{t}^1,\dots,\hat{t}^s)\;
\omega^{(r|s)}\lf(\xx(\st);\frac{\p x}{\p t^1},\dots,\frac{\p x}{\p t^r}\rt|\lf.
     \frac{\p x}{\p \hat{t}^1},\dots,\frac{\p x}{\p \hat{t}^s}\rt),
\end{displaymath}
which does not depend on the parameterization \(\{t^F\}\) of \(N\). 

The second identity is essential\footnote{Strictly speaking Stokes
  theorem would be still valid even without \protect\eq{sform}, but in
  this case \(\dd\omega\) is no longer a differential form and depends
  on second derivatives \(\ddot{\xx}(\st)\).  See
  \protect\cite{ThVor96a} for details.}  for Stokes theorem.  Let now
\(N\) be an \((r+1)|s\)-dimensional submanifold with a boundary \(\p
N\), which is an \(r|s\)-dimensional submanifold. The Stokes theorem
states that
\begin{displaymath}
  \int_N\dd\omega^{(r|s)}=  \int_{\p N}\omega^{(r|s)},
\end{displaymath}
where \(\dd\omega^{(r|s)}\) is an \((r+1)|s\)-form given by 
\begin{multline*}
  [\dd\omega^{(r|s)}](\xx;v_1,v_2,\dots,v_{r+1}|
              \hat{v}_1,\hat{v}_2,\dots,\hat{v}_s)=\\
  (-1)^r v_{r+1}^A\frac{\p \omega^{(r|s)}(\xx;\vv)}{\p
    x^A}-(-1)^{[A][F]}v_F^B\frac{\p^2 \omega(\xx;\vv)}{\p x^B\p v_F^A}.
\end{multline*}

A canonical example of an \(r|s\)-form can be found using a set
\[\cvv=(\cv_1,\cv_2,\dots,\cv_r|\ocv_1,\ocv_2, \dots,\ocv_s)\] of
\(r\) even and \(s\) odd covectors:
\begin{equation}
  \label{canex}
  \omega(\vv)=\ber\lVert\cv_G(v_F)\rVert.
\end{equation}

In the original formulation the \(r|s\)-forms were required to be
nonsingular unless \(\rank(\hat{v}_1,\hat{v}_2, \dots,\hat{v}_s)<s\).
Obviously this is not true for the example given in \eq{canex}, where
the form has a pole whenever one of the odd argument \(\hat v\)
belongs to the null space of of the defining covectors, or 
\(\ocv_1(\hat{v})=\ocv_2(\hat{v})=\cdots=\ocv_s(\hat{v})=0\).

\subsection{Singular $r|s$ forms}
\label{ss:odd}

Equations (\ref{ber}) and (\ref{sform}) have to be satisfied at every
point of the supermanifold and they can be solved independently for
different points.  Let us forget for a moment about the dependence of
\(\omega\) on the point where it is evaluated and concentrate on the
solutions to (\ref{ber}) and (\ref{sform}) at a point.  As a starting
point we will take the solution given by \eq{canex} and produce other
solutions by applying certain operations to it.

Our first operation will map an \(r|s\)-form to an \((r+1)|s\)-form
by taking an exterior product with a covector \(\cv\):
\begin{multline*}
  [\ext_{\cv}\, \omega](v_1,v_2,\dots,v_{r+1}|
   \hat{v}_1,\hat{v}_2,\dots,\hat{v}_r) 
  =\\ (-1)^r\lf[\cv(v_{r+1})-(-1)^{[\cv][F]}
  \cv(v_F)v_{r+1}^A\frac{\p}{\p v^A_F}\rt]\,\omega(\vv). 
\end{multline*}
One can check that if \(\omega\) satisfies eqs.~(\ref{ber})
and~(\ref{sform}), then so does \(\ext_{\cv}\, \omega\).  Exterior
product operators commute (in the super sense) if their parity is
defined as opposite to the parity of the corresponding covector
\begin{displaymath}
  [\ext_{\cv}]=[\cv]+\bar1.
\end{displaymath}
Using even covectors\footnote{We will use different conventions for
  the parity of covectors and \(1\)-forms. Even covector is the one
  which takes even values on even and odd values on odd vectors while
  an odd covector does otherwise. The parity of a \(1\)-form is
  opposite to the parity of the corresponding covector.} we will not
produce new forms by applying \(\ext_{\cv}\) to the canonical form in
(\ref{ber}) since
\begin{equation}
  \label{oddco}
  \ber\lVert\cv_G(v_F)\rVert=\ext_{\cv_1}\ext_{\cv_2}\cdots\ext_{\cv_r}
  (\det\lVert\ocv_i(\hat{v}_j\rVert)^{-1},
\end{equation}
and by applying extra \(\ext_{\cv}\) we will obtain forms of the same
type but with higher even degrees. New forms appear when odd covectors
are used to take an exterior product. The most interesting feature of
these new forms is that their even dimension is not bounded by the
even dimension of the manifold.  Indeed, operators \(\ext_{\ocv_1}\)
and \(\ext_{\ocv_2}\) commute rather than anticommute and therefore
can be iterated indefinitely.  It will be convenient to introduce a
shorter notation for the exterior product
\begin{equation}
  \label{shrte}
  \ext_{\cv}\omega=\cvf\,\omega,
\end{equation}
where \(\cvf\) is the \(1\)-form corresponding to the covector
\(\cv\).  Note that we have to make a distinction between covectors
and \(1\)-forms since the parity of a one form is opposite to the
parity of the corresponding covector.

The second operation that we will define will decrease the degree of
the form by substituting a given vector for one of its arguments.  In
doing so we have two inequivalent choices---to substitute the given
vector for an even or for an odd argument.  Substituting a vector for
an odd argument will change the singularity of the form and operations
that change the singularity will be considered in the next section.
We define the inner product with a vector \(v\) as follows
\begin{displaymath}
  [\inn_v\omega](v_1,v_2,\dots,v_{r-1}|\hat{v}_1,\hat{v}_2,\dots,\hat{v}_s)=
  \omega(v,v_1,v_2,\dots,v_{r-1}|\hat{v}_1,\hat{v}_2,\dots,\hat{v}_s).
\end{displaymath}
We can use odd vectors as well as even in the inner product. Similarly
to the exterior product, inner product operators commute (in the super
sense) if their parity is defined as opposite to the parity of the
corresponding vector
\begin{displaymath}
  [\inn_v]=[v]+\bar1.
\end{displaymath}
Note that inner products with odd vectors can be iterated and
\((\inn_{\hat{v}})^{k}\) is non-zero for any \(k\). Of course, the
iterated operator is only defined on forms with even degree \(r\geq
k\), but it will be useful to extend the space of singular forms by
forms with negative even degree so that \((\inn_{\hat{v}})^{k}\) is
defined on every form for arbitrarily large \(k\).  We will show that
all the form operations like exterior product, differential and Lie
derivative can be defined on the extended space.

Operators of exterior and interior products form together the super
Clifford algebra
\begin{equation}
  \label{cliff}
    [\ext_{\cv_1},\ext_{\cv_2}]=[\inn_{v_1},\inn_{v_2}]=0,\quad
  [\inn_v, \ext_{\cv}] = \cv(v).
\end{equation}

We can describe the space of singular forms with a given singularity
as a free module over the super Clifford algebra (\ref{cliff})
generated from the vacuum state
\(\omega(\vv)=(\det\lVert\ocv_i(\hat{v}_j\rVert)^{-1}\). For even
\(v\) and \(\cv\), \(\inn_v\) is an annihilation operator, 
\begin{displaymath}
  \inn_v\,(\det\lVert\ocv_i(\hat{v}_j\rVert)^{-1}=0,
\end{displaymath}
and \(\ext_{\cv}\) is a creation operator.  For odd \(\ov\) and
\(\ocv\), the situation is a bit more complicated. The vacuum is
annihilated by \(\ext_{\ocv}\) if \(\ocv\) belongs to the space
spanned by \(\ocv_i\) and by \(\inn_{\ov}\) if all \(\ocv_i(\ov)=0\):
\begin{align}
  \label{ezero}
   \ext_{\ocv}\,(\det\lVert\ocv_i(\hat{v}_j)\rVert)^{-1}&=0&&
   \mbox{if \(\ocv=\sum_i a_i\,\ocv_i\)},\\
  \label{izero}
   \inn_{\ov}\,(\det\lVert\ocv_i(\hat{v}_j)\rVert)^{-1}&=0&&
   \mbox{if \(\ocv_i(\ov)=0\) for all \(i=1,\dots,s\)}.
\end{align}

In the two limiting cases, when the odd degree \(s\) of the vacuum
state is either zero or equals the odd dimension \(m\) of the
supermanifold, the even degree, \(r\), of the forms is bounded from
one side.  In the first case the vacuum is annihilated by all
\(\inn_{\ov}\) and therefore \(r\geq0\); in the second case, the
vacuum is annihilated by all \(\ext_{\ocv}\) and the odd degree of the
forms is bounded by the even degree of the supermanifold, \(r\leq n\).
For all the intermediate values of \(s\), \(r\) can be any integer
number.

\subsection{Creating the singularity}
\label{ss:sing}

In the previous section we have shown how to generate a whole space of
differential forms by applying some creation operators to a given
form.  In this construction all the resulting forms had the same odd
dimension and the same singularity as the original one.  Now we want
to introduce operators that will change them both. 

 We define
\begin{multline*}
  [\dl(\inn_{\ov})\omega](v_1,v_2,\dots,v_{r}|
  \hat{v}_1,\hat{v}_2,\dots,\hat{v}_{s-1})=\\
  (-1)^{r}
  \omega(v_1,v_2,\dots,v_{r}|
  \hat{v},\hat{v}_1,\hat{v}_2,\dots,\hat{v}_{s-1}),
\end{multline*}
and
\begin{multline*}
  [\dl(\ext_{\ocv})\omega]
  (v_1,v_2,\dots,v_{r}|\hat{v}_1,\hat{v}_2,\dots,\hat{v}_{s+1})=\\
  \frac{1}{\ocv(\hat{v}_{s+1})}\,
  \omega\lf(\cdots,v_F
  -\frac{\ocv(v_F)\,v_{s+1}}{\ocv(\hat{v}_{s+1})},\cdots\rt),
\end{multline*}
where \(F=1,\dots,r|s\).

The following identities, which easily follow from definitions
partially justify the use of the delta function notation
\begin{displaymath}
  \dl(\inn_{\hat{v}})\,\inn_{\hat{v}}=
  \inn_{\hat{v}}\,\dl(\inn_{\hat{v}})=0,\qquad
 \dl(\ext_{\hat{v}^\vee})\,\ext_{\hat{v}^\vee}=
     \ext_{\hat{v}^\vee}\,\dl(\ext_{\hat{v}^\vee})=0.
\end{displaymath}
It will be useful to define also \(\dl(\inn_{v})\) and
\(\dl(\ext_{\cv})\) for even \(v\) and \(\cv\). In this case
\(\inn_v\) and \(\ext_{\cv}\) are odd operators and it is natural to
assume that \(\dl(\inn_{v})=\inn_v\) and
\(\dl(\ext_{\cv})=\ext_{\cv}\).  Let us now list some of the
commutation properties of the new operators
\begin{align*}
  \dl(\inn_{{u}})\,\dl(\inn_{{v}})&= 
  -\dl(\inn_{{v}})\,\dl(\inn_{{u}})&
  \inn_{u}\,\dl(\inn_{{v}})&= 
  -(-1)^{[u]}\dl(\inn_{{v}})\,\inn_{u}\\
  \dl(\ext_{{u^\vee}})\,\dl(\ext_{{v^\vee}})&= 
  -\dl(\ext_{{v^\vee}})\,\dl(\ext_{{u^\vee}})&
 \ext_{u^\vee}\,\dl(\ext_{{v^\vee}})&= 
  -(-1)^{[u^\vee]}\dl(\ext_{{v^\vee}})\,\ext_{u^\vee}.
\end{align*}
This shows that all these operators will become supercommutative if we
assign odd parity to \(\dl(\inn_{v})\) and \(\ext_{{v^\vee}}\) no
matter what parity \(v\) and \(\cv\) have.
\begin{displaymath}
  [\dl(\inn_{v})]=[\dl(\ext_{\cv})]=\bar1.
\end{displaymath}
Using \(\dl(\ext_{\ocv})\) one can generate all odd vacuum forms from
\(\omega^{(0|0)}=1\). Indeed, one can show that 
\begin{displaymath}
  (\det\lVert\ocv_i(\ov_j)\rVert)^{-1}=
  \dl(\ext_{\ocv_1})\,\dl(\ext_{\ocv_2})\,\cdots\,\dl(\ext_{\ocv_s})\,1,
\end{displaymath}
or extending the convention of \eq{shrte}, we can write
\begin{displaymath}
   (\det\lVert\ocv_i(\ov_j)\rVert)^{-1}=
  \dl(\Tilde{\Hat{v}}_1^\vee)\,\dl(\Tilde{\Hat{v}}_2^\vee)\cdots  
  \dl(\Tilde{\Hat{v}}_s^\vee).
\end{displaymath}
Furthermore, using the convention that the delta function is identity
on the odd arguments, we can rewrite the form in \eq{oddco} as
\begin{displaymath}
  \omega(\vv)=\ber\lVert\cv_G(v_F)\rVert
       =\prod_{G=1}^{r|s}\dl(\Tilde{v}_G^\vee).
\end{displaymath}

\subsection{Commutation relations}
\label{ss:comm}

As it was stated above, operators \(\ext_{\cv}\) and their delta
functions super commute and so do \(\inn_{v}\) and \(\dl(\inn_v)\).
Now we will analyze their mixed products and the products involving
two delta functions.

As it is the case with ordinary delta functions the product of two
delta symbols with the same argument leads to divergences.  On the
other hand if arguments are linearly independent then the compositions
of the corresponding operators are finite and satisfy the following
identities
\begin{align}
  \label{dldl}
  \dl(\inn_{\ov_1+\ov_2})\,\dl(\inn_{\ov_2})
              &=\dl(\inn_{\ov_1})\,\dl(\inn_{\ov_2})&
  \dl(\ext_{\ocv_1+\ocv_2})\,\dl(\ext_{\ocv_2})
              &=\dl(\ext_{\ocv_1})\,\dl(\ext_{\ocv_2}).
\end{align}

Operators \(\dl(\ext_{\ocv})\) and \(\dl(\inn_{\ov})\) do not satisfy
the same commutation relations as \(\ext_{\ocv}\) and \(\inn_{\ov}\),
in fact their commutators cannot even be defined except for the
trivial case \(\ocv(\ov)=0\) when they commute.  However, the
following weaker form of the relation
\([\ext_{\cv},\inn_{\ov}]=\ocv(\ov)\), which states that
\begin{align*}
  \ext_{\ocv}\,\inn_{\ov}\,\omega&=\ocv(\ov)\,\omega&&
  \mbox{if \(\ext_{\ocv}\,\omega=0\)}\\ \noalign{and}
  \inn_{\ov}\,\ext_{\cv}\,\omega&=\ocv(\ov)\,\omega&&
  \mbox{if \(\inn_{\ov}\,\omega=0\)}
\end{align*}
has an analog for  \(\dl(\ext_{\ocv})\) and \(\dl(\inn_{\ov})\):
\begin{align}
  \label{dledli}
  \dl(\ext_{\ocv})\,\dl(\inn_{\ov})\,\omega&=\frac{1}{\ocv(\ov)}\,\omega&&
  \mbox{if \(\ext_{\ocv}\,\omega=0\)}\\ \noalign{and}
  \label{dlidle}
  \dl(\inn_{\ov})\,\dl(\ext_{\cv})\,\omega&=\frac{1}{\ocv(\ov)}\,\omega&&
  \mbox{if \(\inn_{\ov}\,\omega=0\)}.
\end{align}
The appearance of \(\ocv(\ov)\) in the denominator is consistent with
the fact that delta function is homogeneous of degree \(-1\).

It follows easily from the definitions that the (anti)commutators of
\(\ext_{\ocv}\) with \(\dl(\inn_{\hat{v}})\) and \(\inn_{\hat{v}}\)
with \(\dl(\ext_{\ocv})\) are both proportional to \(\ocv(\ov)\) with
the operator factor depending only on the argument of the delta
function.  This observation allows us to define new operators which we
will denote by \(\dl'(\inn_{\hat{v}})\) and \(\dl'(\ext_{\ocv})\)
\begin{align}
  \label{dlprime}
  [\ext_{\ocv},\dl(\inn_{\hat{v}})]&=\ocv(\hat{v})\,\dl'(\inn_{\hat{v}})&
  [\inn_{\hat{v}},\dl(\ext_{\ocv})]&=\ocv(\hat{v})\,\dl'(\ext_{\ocv}),
\end{align}
repeating the logic we can define operators with multiple derivatives
of delta functions
\begin{align*}
  [\ext_{\ocv},\dl^{(n)}(\inn_{\hat{v}})]&=\ocv(v)\,\dl^{(n+1)}(\inn_{\hat{v}})&
  [\inn_{\hat{v}},\dl^{(n)}(\ext_{\ocv})]&=\ocv(v)\, \dl^{(n+1)}(\ext_{\ocv}).
\end{align*}
The degrees of the resulting operators are given by
\begin{align*}
  \deg(\dl^{(n)}(\inn_{\hat{v}}))&=n|(-1)&
  \deg(\dl^{(n)}(\ext_{\ocv}))&=(-n)|1.
\end{align*}
Recall that for even \(v\) and \(\cv\), and therefore odd \(\inn_v\)
and \(\ext_{\cv}\) the delta functions of the operators were
identified with the operators themselves.  Identities in
(\ref{dlprime}) would still hold under this convention since
\(\dl'(\inn_v)=\dl'(\ext_{\cv})=1\) and both identities would become
equivalent to the last identity in (\ref{cliff}).

We will also use derivatives of delta functions to write down the
singular forms without explicit use of \(\inn_{\ov}\) operators.
Recall that the space of singular forms was defined as a module
generated by all possible \(\ext_{\cv}\) and \(\inn_{v}\) operators
from a vacuum \(0|s\)-form which can be written as
\(\dl(\Tilde{\Hat{v}}^\vee_1)\cdots\dl(\Tilde{\Hat{v}}^\vee_s)\).
Each application of \(\inn_{\ov}\) results in a derivative of a delta
function
\begin{displaymath}
  \inn_{\ov}\dl(\Tilde{\Hat{v}}^\vee_1)\cdots\dl(\Tilde{\Hat{v}}^\vee_s)=
  \sum_{k=1}^s\ocv(\ov)\,\dl(\Tilde{\Hat{v}}^\vee_1)\cdots
              \dl'(\Tilde{\Hat{v}}^\vee_k)\cdots
              \dl(\Tilde{\Hat{v}}^\vee_s).
\end{displaymath}
One of the benefits of this new notation is that \eq{izero} becomes a
tautology.

\subsection{The differential and the Lie derivative}
\label{ss:dnlie}

We have briefly introduced the de Rham differential in
section~\ref{ss:odd}.  Let us repeat the definition.  The de Rham
differential, \(\dd\), is a nilpotent differential operator which maps
\(\Omega^{r|s}(M)\) to \(\Omega^{(r+1)|s}(M)\) and can be written in
coordinates as
\begin{multline*}
  [\dd\omega](\xx;v_1,v_2,\dots,v_{r+1}|
              \hat{v}_1,\hat{v}_2,\dots,\hat{v}_s)=\\
  (-1)^r v_{r+1}^A\frac{\p \omega(\xx;\vv)}{\p
    x^A}-(-1)^{[A][F]}v_F^B\frac{\p^2 \omega(\xx;\vv)}{\p x^B\p v_F^A}.
\end{multline*}
Another important differential operation on the differential forms is
the Lie derivative along a vector field.  The Lie derivative can be
defined as usual by differentiating the drag of a form produced by the
vector field and this would produce the following explicit formula:
\begin{displaymath}
    \Lie_v\omega(\xx;\vv)=v^A\frac{\p\omega(\xx;\vv)}{\p x^A}
        +(-1)^{[F][v]}\,v^B_F\,
  \frac{\p v^A}{\p x^B}\frac{\p\omega(\xx;\vv)}{\p v^A_F}.
\end{displaymath}

In a complete analogy with the purely even case, the differential and
the Lie derivative satisfy the following super commutator relations
with the inner product:
\begin{align}
  \label{lieinn}
  [\Lie_v,\inn_u]=\inn_{[v,u]},\\ \noalign{and}
  \Lie_v=[\dd,\inn_v].
\end{align}
These identities make it possible to extend the definition of \(\dd\)
and \(\Lie_v\) to the differential forms of negative degrees which can
be generated by repeated application of \(\inn_{\ov}\) with odd
\(\ov\).

For our applications we will also need the commutation relations of
\(\dd\) and \(\Lie_v\) with the derivatives of delta function of the
inner product operator
\begin{align}
  \label{liedl}
  [\Lie_{\hat{v}},\dl^{(n)}(\inn_{\hat{v}})]&=
  \dl^{(n+1)}(\inn_{\hat{v}})\,\inn_{[\hat{v},\hat{v}]},
\\
\label{ddli}
  {}[\dd,\dl^{(n)}(\inn_{\hat{v}})]&=
  \dl^{(n+1)}(\inn_{\hat{v}})\,\Lie_{\hat{v}}+\frac{1}{2}
  \dl^{(n+2)}(\inn_{\hat{v}})\,\inn_{[\hat{v},\hat{v}]},
\end{align}
where \([\hat{v},\hat{v}]=2\,\hat{v}^2\) is the super Lie bracket
(anticommutator) of the odd vector field \(\hat{v}\) with itself. Note
that the all the commutation relations involving delta functions would
have exactly the same form if the delta function was replaced with an
analytical function. 

\subsection{Picture changing operators}
\label{ss:pco}

Let \(\hat{v}\) be a non-vanishing odd vector field on the
supermanifold.  We define the picture changing operator associated
with \(\hat{v}\) as follows
\begin{equation}
  \label{pcodf}
  \pco_{\hat{v}}=\frac{1}{2}\big(\dl(\inn_{\hat{v}})\,\Lie_{\hat{v}}
       -\Lie_{\hat{v}}\,\dl(\inn_{\hat{v}})\big)=
  \dl(\inn_{\hat{v}})\,\Lie_{\hat{v}}+\dl'(\inn_{\hat{v}})\,\inn_{\hat{v}^2}.
\end{equation}
The picture changing operator commutes with the de Rham differential
\begin{displaymath}
   [\dd,\pco_{\hat{v}}]=0.
\end{displaymath}
This property of \(\pco_{\hat{v}}\) easily follows from the
commutation relations (\ref{liedl}), (\ref{ddli}) and (\ref{lieinn}), but
it is useful to give the following empirical reason.  The picture
changing operator can be formally written as 
\begin{equation}
  \label{pcofrml}
  \pco_{\hat{v}}=[\dd,\Theta(\inn_{\hat{v}})],
\end{equation}
where \(\Theta(x)\) is the Heavyside step function
\(\Theta'(x)=\dl(x)\).  Since \(\dd^2=0\) the vanishing of the
commutator \([\dd,\pco_{\hat{v}}]\) trivially follows from
\eq{pcofrml}.  Note that although \eq{pcofrml} expresses
\(\pco_{\hat{v}}\) as a \(\dd\)-exact operator, this does not mean
that it would vanish in cohomology because \(\Theta(\inn_{\hat{v}})\)
cannot be defined as an operator within the de Rham complex.

Since picture changing operators contain \(\dl(\inn_{\ov})\), they
cannot be composed unless their arguments are linearly independent at
every point of the supermanifold. For two such linearly independent
vector fields, \(\ov_1\) and \(\ov_2\), it is interesting two find the
commutator of their corresponding picture changing operators.  It turns
out that the commutator can be written as
\begin{equation}
  \label{pcocmt}
  [\pco_{\ov_1},\pco_{\ov_2}]=
    [\dd,\dl(\inn_{\ov_1})\,\dl(\inn_{\ov_2})\inn_{[\ov_1,\ov_2]}],
\end{equation}
which means that on the cohomology, picture changing operators form a
commutative algebra. 

It is natural to define the higher order picture changing operators as
symmetrized products 
\begin{displaymath}
  \pco_{\ov_1\cdots\ov_n}^{(n)}=\sum_{\sigma\in\mathfrak{S}_n}
  \pco_{\ov_{\sigma(1)}}\cdots\pco_{\ov_{\sigma(n)}}.
\end{displaymath}
It follows from \eq{pcocmt} that
\begin{equation}
  \label{pconp}
  \pco_{\ov_1\cdots\ov_{n+1}}^{(n+1)}=\pco_{\ov_{n+1}}\,
      \pco_{\ov_1\cdots\ov_n}^{(n)}+[\dd,\cdots].
\end{equation}

\subsection{A toy problem: $\Omega^{\cdot|\cdot}(S^{1|1})$}
\label{ss:s11}

In order to illustrate the properties of the picture changing
operators, let us analyze the de Rham complex of the simplest compact
supermanifold---the supercircle.  In fact there are two nonequivalent
compact supermanifolds of dimension \(1|1\), one with a periodic odd
coordinate and the other with an antiperiodic one, but the analysis
will be exactly the same for both cases. Let \(x\) and \(\xi\) be the
coordinates on \(S^{1|1}\) such that \((x+2\,\pi,\xi)=(x,\pm\xi)\).
Since the odd dimension of \(S^{1|1}\) is one, we have only two
pictures: \(s=0\) and \(s=1\). The de Rham complex in picture \(s=0\)
is infinite to the right (\(r\geq0\)) and in picture \(s=1\) it is
infinite to the left \((r\leq1)\).  This situation is depicted in the
following diagram


\begin{equation}
  \label{scmplx}
  \begin{CD}
    @. \fbox{\(\scriptstyle0\)} @. \fbox{\(\scriptstyle1\)}
  @. \fbox{\(\scriptstyle2\)} @. \fbox{\(\scriptstyle2\)}@. \cdots\\
    @.  0@>\dd>>\Omega^{0|0}@>\dd>>\Omega^{1|0}@>\dd>>\Omega^{2|0}@>\dd>>\cdots\\
    @. @.  @AA{\pco}A @AA{\pco}A\\
    \cdots@>\dd>>\Omega^{(-1)|1}@>\dd>>\Omega^{0|1}@>\dd>>\Omega^{1|1}@>\dd>>0 @. \\
   \cdots   @. \fbox{\(\scriptstyle2\)} @. \fbox{\(\scriptstyle2\)} @.
   \fbox{\(\scriptstyle1\)} @.  \fbox{\(\scriptstyle0\)} @.
  \end{CD}
\end{equation}

Differential forms in picture \(s=0\) can be written as polynomials in
\(\d x\) and \(\d\xi\), so the space \(\Omega^{r|0}\) is spanned by
\begin{align*}
  \omega^{(r|0)}_1&=\d x(\d\xi)^{r-1}f_1(x,\xi)\\
  \noalign{and}
  \omega^{(r|0)}_2&=(\d\xi)^{r}f_2(x,\xi).
\end{align*}
Similarly \(\Omega^{r|1}\) is spanned by 
\begin{align*}
  \omega^{(r|1)}_1&=\d x\dl^{(1-r)}(\d\xi)g_1(x,\xi)\\
  \noalign{and}
  \omega^{(r|1)}_2&=\dl^{(-r)}(\d\xi)g_2(x,\xi).
\end{align*}
This shows that the dimensions of \(\Omega^{r|s}(S^{1|1})\) at a point
are as given by boxed numbers in the diagram (\ref{scmplx}).  One can
easily calculate the differential of these forms and find that the
only nonzero cohomology classes can be represented by \(1\), \(\d x\),
\(\xi\,\dl(\d\xi)\) and \(\d x\,\dl(\d\xi)\).
\begin{displaymath}
  H^{r|s}(S^{1|1})=\begin{cases}
    \IR& \text{for \(r=0,1\) and \(s=0,1\)}\\
    0& \text{otherwise}
  \end{cases}
\end{displaymath}
Note that it is only due to the presence of negative degree forms that
the closed \(0|1\) form \(\dl(\d\xi)\) is trivial and the cohomology
is the same in the two pictures:
\begin{displaymath}
  \dl(\d\xi)=-\dd(\xi\,\dl'(\d\xi)).
\end{displaymath}

It follows easily from the dimensional analysis that picture changing
operator is not invertible.  Nevertheless it is possible to define an
operator \(\ipco\) which is an inverse of \(\pco\) restricted to a
subspace of \(\Omega^{\cdot|\cdot}(S^{1|1})\).  Moreover, the pair
\(\pco\) and \(\ipco\) induces an isomorphism of the \emph{cohomology}
in the two pictures.

According to Shander's theorem \cite{shander} any odd vector field is
locally equivalent to one of the following two up to a change of
coordinates
\begin{align*}
  \p_{\xi}&=\frac{\p}{\p\xi},\\
  \noalign{and}
  D&=\frac{\p}{\p\xi}+\xi\,\frac{\p}{\p x}.
\end{align*}
The second vector field is more interesting in a way that its square
is not zero
\begin{displaymath}
  D^2=\frac{\p}{\p x}.
\end{displaymath}
Let us start with \(\pco_{\p_\xi}\).  A simple calculation shows that
\begin{displaymath}
  \pco_{\p_\xi}\xi\,\dl(\d\xi)\,f(x,\xi)=f(x,\xi)\quad\text{and}
  \quad
  \pco_{\p_\xi}\d x\,\dl(\d\xi)\,f(x,\xi)=\d x\,\p_\xi f(x,\xi).
\end{displaymath}
The inverse picture changing operator is therefore
\begin{displaymath}
  \ipco_{\p_\xi}=\xi\,\dl(\ext_{\d\xi}).
\end{displaymath}
Indeed,
\begin{displaymath}
  \ipco_{\p_\xi}f(x,\xi)=\xi\,\dl(\d\xi)\,f(x,\xi)\quad\text{and}
  \quad
  \ipco_{\p_\xi}\d x\,\p_\xi f(x,\xi)=\d x\,\dl(\d\xi)\,f(x,\xi).
\end{displaymath}
Similar analysis applies to \(\pco_D\) only with slightly more
complicated formulae.  The inverse picture changing operator in this
case is given by 
\begin{equation}
  \label{ipcod}
  \ipco_D=\xi\,\dl(\ext_{\d\xi})+\ext_{\d z}\dl'(\ext_{\d\xi}),
\end{equation}
which follows from the identities
\begin{align*}
  \ipco_D\,f(x,\xi)&
     =(\xi\,\dl(\d\xi)+\d z\dl'(\d\xi)))\,f(x,\xi),\\
  \pco_D\,(\xi\,\dl(\d\xi)+\d z\dl'(\d\xi)))\,f(x,\xi)&
     =f(x,\xi),\\
  \ipco_D\,(\d\xi\,f(x,\xi)+(\d x-\xi\,\d\xi)\,Df(x,\xi))&
     =\d x\dl(\d\xi)\,f(x,\xi),\\
  \pco_D\,\d x\dl(\d\xi)\,f(x,\xi)&
     =\d\xi\,f(x,\xi)+(\d x-\xi\,\d\xi)\,Df(x,\xi).
\end{align*}
The product of \(\pco\) with \(\ipco\) as operators in
\(\Omega^{\cdot|\cdot}\) is not the unit operator but
a projector.  However, these operators induce the following maps in
cohomology 
\begin{align*}
  \hpco:&\;H^{\cdot|1}\to H^{\cdot|0},\\
  \ihpco:&\;H^{\cdot|0}\to H^{\cdot|1},
\end{align*}
were the second map is the true inverse of the first
\begin{align*}
   \hpco\circ\ihpco&=\mathbbm{1}\quad\text{on \(H^{\cdot|0}\)}\\
   \noalign{and}
   \ihpco\circ\hpco&=\mathbbm{1}\quad\text{on \(H^{\cdot|1}\)}
\end{align*}

\section{Superstring picture changing operators}
\setcounter{equation}{0}
\label{sec:sstring}
In this section we will describe the picture changing operators as
they appear in string theory.  The main idea is to show that the
supersting picture changing operators are closely related to the
similar operators on the super de~Rham complex.  Picture changing
operators in supersrings were extensively studied by different
methods, and even though their geometrical significance was not fully
realized, most of the results that we will present here can be found
in the literature
\cite{Witten86,Yamron87,en87,n-q88,amn88,am88,am90,amz90,DPol95}. 

\subsection{A review of the FMS approach}
\label{ss:review}

\subsubsection{Superconformal ghosts}
Superconformal gauge fixing in the string functional integral produces
two pairs of ghost fields: fermionic \(b\) and \(c\) of conformal
dimensions \(\Delta_b=2\) and \(\Delta_c=-1\) and bosonic \(\bt\) and
\(\gm\) of conformal dimensions \(\Delta_\bt=3/2\) and
\(\Delta_\gm=-1/2\).  Their operator products are given by 
\begin{align*}
  c(z)b(w)&=\frac{1}{z-w}+\no{c\,b}(w)+\cdots\\
  \noalign{and}
  \gm(z)\bt(w)&=\frac{1}{z-w}+\no{\gm\,\bt}(w)+\cdots;
\end{align*}
and the energy-momentum tensors are
\begin{align}
  \label{emtbc}
  L_{bc}&=2\no{\p c\,b}+\no{c\,\p b}\\
  L_{\bt\gm}&=-\frac{1}{2}\no{\gm\,\p\bt}-\frac{3}{2}\no{\p\gm\,\bt}.
\end{align}
The central charges of the corresponding Virasoro representations are
\(c_{bc}=-26\) and \(c_{\bt\gm}=11\).  In addition to the
energy-momentum tensor \(L_{\rm gh}=L_{bc}+L_{\bt\gm}\), which is the
variation of the ghost action w.r.t.\ the graviton field, we need to
introduce the supercurrent \(G_{\rm gh}\) given by the variation of
the ghost action w.r.t.\ the gravitino
\begin{displaymath}
  G_{\rm gh}=-2\no{c\,\p\bt}-3\no{\p c\,\bt}+\no{\gm\, b}
\end{displaymath}
Operators \(L_{\rm gh}\) and \(G_{\rm gh}\) together generate a super
Virasoro algebra with the central charge
\(\hat{c}=(2/3)(c_{bc}+c_{\bt\gm})=-10\).

\subsubsection{Bosonization of superconformal ghosts}

Fermionic ghosts \(b\) and \(c\) can be bosonized using the following
procedure.  One has to introduce a bosonic field \(\sigma\) such that
\begin{equation}
  \label{jsig}
  j_\sigma=\p\sigma=-\no{b\,c}.
\end{equation}
The energy momentum tensor \(L_{bc}\) can be rewritten in term of
\(j_\sigma\) as
\begin{equation}
  \label{emts}
  L_{bc}=\frac{1}{2}\no{j_\sigma^2}+3\,\p j_\sigma,
\end{equation}
which can be proven by a direct substitution of the r.h.s.\ of
\eq{jsig} and comparison with \eq{emtbc}.  Fermionic fields \(b\) and
\(c\) can be recovered back from \(\sigma\) as
\begin{align*}
  c&=e^{\sigma}&  b&=e^{-\sigma}.
\end{align*}

If we proceed similarly for the bosonic \(\bt\) and \(\gm\) and,
following the sign convention of \cite{fms86}, introduce
\begin{displaymath}
  j_\phi=-\p\phi=-\no{\bt\,\gm}
\end{displaymath}
the energy momentum tensor constructed out of \(j_\phi\),
\begin{equation}
  \label{emtphi}
  L_{\phi}=-\frac{1}{2}\no{j_\phi^2}+\p j_\phi,
\end{equation}
would not reproduce \(L_{\bt\gm}\) because of an extra term which
comes out of \(\no{j_\phi^2}\)
\begin{displaymath}
   \no{j_\phi^2}=\no{\bt\,\p\gm}-\no{\p\bt\,\gm}
              +\frac{1}{2}(\no{\bt^2\,\gm^2}+\no{\gm^2\,\bt^2}).
\end{displaymath}
Note that if \(\bt\) and \(\gm\) were fermions, this term would be
zero as it is for \(b\,c\) ghosts.  Therefore instead of \eq{emts}
we now have
\begin{displaymath}
  L_{\bt\gm}=L_{\phi}-\frac{1}{4}(\no{\bt^2\,\gm^2}+\no{\gm^2\,\bt^2}).
\end{displaymath}
Furthermore, the exponents \(e^\phi\) and \(e^{-\phi}\) cannot
reproduce \(\bt\) and \(\gm\) since they are fermions and their
operator product expansion is
\begin{equation}
  \label{dlope}
  \no{e^{\phi(z)}}\no{e^{-\phi(w)}}=(z-w)\,\no{e^{\phi(z)-\phi(w)}}
  =(z-w)+(z-w)^2\p\phi+\cdots.
\end{equation}
It turns out that \(e^\phi\) and \(e^{-\phi}\) correspond to the delta
functions of the bosonic ghosts
\begin{align*}
  \dl(\bt)&=e^\phi&     \dl(\gm)&=e^{-\phi}.   
\end{align*}
In order to find the bosonization formulae for \(\bt\) and \(\gm\) one
has to introduce an auxiliary system of free fermions \(\eta\) and
\(\xi\) with conformal weights \(1\) and \(0\) such that
\begin{align*}
  \eta&=\p\gm\,\dl(\gm)&   \p\xi=\p\bt\,\dl(\bt).
\end{align*}
The extra term in \eq{emtphi} can be identified with the
energy-momentum tensor of the \(\eta\,\xi\) system:
\begin{displaymath}
  L_{\eta\xi}=-\eta\p\xi=-\frac{1}{4}(\no{\bt^2\,\gm^2}+\no{\gm^2\,\bt^2}).
\end{displaymath}
The auxiliary fermions can be bosonized in the same manner as the
\(b\,c\) ghosts using 
\begin{align*}
  j_\chi&=\p\chi=-\no{\xi\,\eta}& \xi&=e^\chi&   \eta&=e^{-\chi}.
\end{align*}
The bosonic ghosts can now be recovered through the following formulae
\begin{align*}
  \bt&=e^{-\phi}\p\xi=e^{-\phi+\chi}\p\chi&
  \gm&=e^{\phi}\eta=e^{\phi-\chi}
\end{align*}
Bosonization formulae are extremely useful in calculations, but they
tend to hide the geometrical meaning of the superconformal ghosts.
Since exposing the geometry is our primary goal in this paper, we will
use the bosonization formulae only to show the relation between the
new geometric formalism and the original approach.

\subsubsection{The BRST operator}
When the superconformal ghosts are combined with a \(\hat c=10\)
matter representation of the super Virasoro algebra \(L_{\rm m}\),
\(G_{\rm m}\), one can define a nilpotent BRST operator as
\begin{equation}
  \label{brstdf}
  Q= Q^{(0)}+Q^{(1)}+Q^{(2)}
\end{equation}
where
\begin{align*}
  Q^{(0)}&=\oint\frac{\d z}{2\pi i}
       \no{c(z)\,(L_{\rm m}(z)+L_{\bt\gm}(z)+\frac{1}{2}L_{bc}(z))}\\
  Q^{(1)}&=-\oint\frac{\d z}{2\pi i}\gm(z)\,G_{\rm m}(z)\\
  Q^{(2)}&=-\oint\frac{\d z}{2\pi i}\gm^2(z)\,b(z)
\end{align*}
The first term \(Q^{(0)}\) has the same structure as the BRST operator
of the bosonic string in the background described  by  the conformal
field theory with the energy momentum tensor given by \(L=L_{\rm
  m}+L_{\bt\gm}\). 
\subsubsection{Picture changing operation}
Bosonization formulae for the \(\bt\gm\) system do not contain the
zero mode of the auxiliary fermion \(\xi\).  Therefore a state like
\(\xi_0\ket\psi\) does not belong to the superstring BRST
complex. This fact was used by FMS to introduce their picture changing
operation. Let \(\ket\psi\) be a BRST closed state, then the state 
\begin{displaymath}
  \pco\ket\psi=Q\,\xi_0\ket\psi,
\end{displaymath}
will not contain \(\xi_0\) and will be BRST closed simply because
\(Q^2=0\).  Although the definition of \(\pco\ket\psi\) suggests that
it is BRST trivial, it is not since \(\xi_0\ket\psi\) does not belong
to the BRST complex. The picture changing operator was first
introduced in \cite{ver87} as
\begin{displaymath}
  \pco(z)=[Q,\xi(z)].
\end{displaymath}
One can think of \(\pco(z)\) as a conformal field corresponding to the
state \(\pco\ket0\equiv\pco(0)\ket0\).

\subsection{New geometrical approach}
\subsubsection{Superconformal ghosts as semi-infinite forms}
\label{ss:ghforms}

In our recent work \cite{bel97}, we have shown that the ghost sector
of the superstring can be interpreted as a space of semi-infinite
singular forms on the algebra of superconformal vector fields on
\(S^{1|1}\). Let \(l(z)\) and \(g(z)\) be the even and odd vector
fields concentrated\footnote{The delta function in the following
  formulae are should be interpreted as distributions on the unit
  circle and properly normalized.} at the point \(z\)
\begin{align*}
  l(z)&=\dl(z'-z)\frac{\p}{\p
    z'}=\sum_{n=-\infty}^\infty\frac{l_n}{z^{n+2}},\\
  g(z)&=\dl(z'-z)\frac{\p}{\p
    z'}=\sum_{r=-\infty}^\infty\frac{g_r}{z^{r+\frac{3}{2}}}.
\end{align*}
The conjugate covectors we will denote by \(l^\vee(z)\) and
\(g^\vee(z)\) and define by
\begin{align*}
  \la l^\vee(z),l(w)\ra&=\dl(z-w)&  \la g^\vee(z),g(w)\ra&=\dl(z-w).
\end{align*}
Conformal fields of the superconformal ghosts can now be written as
inner and exterior product operators.\footnote{Here we use a
  normalization for the bosonic ghosts which is slightly different form
  that of Friedan, Martinec and Shenker \cite{fms86}. Our \(\bt\) is
  \(2\,\bt\) in the FMS notation and our \(\gm\) is \(\frac{1}{2}\gm\).}  
\begin{align}
  \label{bbdf}
  b(z)&=\inn_{l(z)}&  \bt(z)&=\inn_{g(z)}\\
  \label{cgdf}
  c(z)&=\ext_{l^\vee(z)}&\gm(z)&=\ext_{g^\vee(z)}
\end{align}
The operator product expansions for \(b\), \(c\), \(\bt\), \(\gm\),
\(\dl(\bt)\) and \(\dl(\gm)\) now follow easily from the properties of
\(\inn\), \(\ext\) and delta functions of them. For example
eqs.~(\ref{dledli}) and ~(\ref{dledli}) translate into (\emph{c.f.}
\eq{dlope})
\begin{displaymath}
  \dl(\bt(z))\dl(\gm(w))=(z-w)-(z-w)^2\no{\bt\,\gm}(w)+\cdots
\end{displaymath}
and eqs.~(\ref{dldl})---to
\begin{align*}
  \dl(\bt(z))\dl(\bt(w))&=\frac{1}{z-w}\no{\dl(\p\bt)\dl(\bt)}(w)+
  \frac{1}{2}\no{\p^2\bt\dl'(\p\bt)\dl(\bt)}(w)+\cdots\\
  \noalign{and}
  \dl(\gm(z))\dl(\gm(w))&=\frac{1}{z-w}\no{\dl(\p\gm)\dl(\gm)}(w)+
  \frac{1}{2}\no{\p^2\gm\dl'(\p\gm)\dl(\gm)}(w)+\cdots&
\end{align*}
The ghost vacua can be identified with the semi-infinite forms using
the identity
\begin{equation}
  \label{gh-sif}
  \exp(r\,\sigma-s\,\phi)\ket0=\ket{r|s},
\end{equation}
where \(\ket{r|s}\) is a semi-infinite form of degree \(r|s\) given by
\begin{displaymath}
  \ket{r|s}=\tilde{l}^\vee_{-r+2}\,\dl(\tilde{g}^\vee_{-s+\frac{3}{2}})\,
    \tilde{l}^\vee_{-r+3}\,\dl(\tilde{g}^\vee_{-s+\frac{5}{2}})\,\cdots.
\end{displaymath}
Together with eqs.~(\ref{bbdf}) and~(\ref{cgdf}), this identity
provides the translation between the language of FMS and the language
of semi-infinite forms.

\subsubsection{The BRST complex}
In order for the differential which is also known as the BRST operator
\(Q\) to be nilpotent the semi-infinite forms should take values in a
matter representation of the super Virasoro algebra with the central
charge \(\hat c=10\).  Let \(L_{\text{m}}(z)\) and \(G_{\text{m}}(z)\)
be the super Virasoro generators in the matter representation, then we
can write the Lie derivative operator on the semi-infinite forms as
\begin{align*}
  \Lie_{l(z)}&=L(z)=L_{\rm gh}(z)+L_{\rm m}(z)\\
  \Lie_{g(z)}&=G(z)=G_{\rm gh}(z)+G_{\rm m}(z)
\end{align*}
The BRST operator appears a differential for these semi-infinite forms
and coincides with the one defined in \eq{brstdf}.

\subsubsection{Picture changing operator(s)}
\label{sss:pco}

Picture changing operation on the superstring BRST complex can be
introduced in two ways: using the language of semi-infinite forms or
the language of conformal fields.  The two approaches are equivalent
and they are related by the usual CFT correspondence of states and
conformal fields.  In the language of semi-infinite forms, one can
associate a picture changing operator with every odd generator \(g_r\)
of the super Virasoro algebra using \eq{pcodf}
\begin{displaymath}
  \pco_{g_r}=\frac{1}{2}(G_r\,\dl(\bt_r)-\dl(\bt_r)\,G_r)
   =\dl(\bt_r)\,G_r+\dl'(\bt_r)\,b_{2r}
\end{displaymath}
Alternatively one can introduce one conformal field of conformal
dimension zero
\begin{displaymath}
  \pco(z)=\pco_{g(z)}=\frac{1}{2}\big(
  \no{G\,\dl(\bt)}(z)-\no{\dl(\bt)\,G}(z)\big)
  =\no{\dl(\bt)\,G}(z)+\no{\dl'(\bt)\,\p b}(z).
\end{displaymath}
Just like the picture changing operator of the singular de Rham
complex, the conformal field \(\pco\) can be formally written as a
commutator with the differential 
\begin{displaymath}
  \pco(z)=[Q,\Theta(\bt(z))],
\end{displaymath}
where \(\Theta(\bt(z))\) is a fermionic field of conformal dimension
zero. In the FMS formalism this field can be identified with the
auxiliary fermion
\begin{displaymath}
  \xi(z)=\Theta(\bt(z)).
\end{displaymath}
This was first pointed out in \cite{ver87}.

The picture changing operator acts differently on different vacua. One
can easily show that
\begin{displaymath}
  \lim_{z\to0}\pco(z)\ket{r|s}=\pco_{g_{-s+\frac{3}{2}}}\ket{r|s}.
\end{displaymath}
This allows us to apply \(\pco(z)\) repeatedly without having a
picture changing operator with the same argument appear twice.

It was shown in \cite{lz92,lz95} that although the normal ordered
product is neither commutative nor associative it generates a
supercommutative and associative product in the cohomology called the
dot product. Using the normal ordered product we can define the higher
order picture changing operators as
\begin{displaymath}
  \pco^{(n)}=\no{\pco^n}.
\end{displaymath}
The on-shell associativity of the normal ordered product immediately
shows that
\begin{equation}
  \label{pcosn}
  \pco^{(n+1)}=\no{\pco\,\pco^{(n)}}+[Q,\cdots].
\end{equation}
This equation is the string version of \eq{pconp}.  Note that although
the properties of the dot product make it trivial to prove \eq{pcosn},
they do not help in determining the BRST trivial part.  This part can
be easily deduced from \eq{pcocmt} or computed in a brute force
calculation as in \cite{DPol95}.\footnote{Note that in the work
  \cite{DPol95} a different definition of the normal ordering is
  implicitely used.  The normal ordering of \cite{DPol95} is the
  symmetric normal ordering and the picture changing operator appears
  without the extra \(\dl'(\bt)\p b\) term.  It is not clear though
  what prescription should be made there for the multiple normal
  ordered products.}

\subsubsection{Inverse picture changing operator}
An inverse picture changing operator \(Y\) was discovered in the
bosonized form by the authors of \cite{fms86} and further described in
\cite{Witten86}. An explicit expression for \(Y\) in terms of ghost
fields was deduced in \cite{Yamron87} to have the following simple
form
\begin{equation}
  \label{sipcodf}
  \ipco(z)=\no{c(z)\,\dl'(\bt(z))}.
\end{equation}
This expression resembles the second term in \eq{ipcod}, but a closer
look reveals that it is analogous to the whole expression for \(Y_D\).
To make the analogy complete we have to relate \(\bt\) to \(\inn_D\)
and \(c\) to \(\d x-\xi\,\d\xi\). Eq.~(\ref{ipcod}) can be
rewritten as
\begin{displaymath}
  \ipco_D=(\d x-\xi\,\d\xi)\dl'(\d\xi),
\end{displaymath}
which looks very similar to \eq{sipcodf}.  This analogy is not
accidental as will be explained in section~\ref{sec:conn}.

In the case of supermanifolds we were able to construct the inverse
picture changing operator only for the \(1|1\)-dimensional case.
Generalization to the higher dimensions would require a better
understanding of the global structure of the supermanifold and may
even not be possible.  In the case of the superstring we can take
advantage of the additional algebraic structure on the BRST cohomology
supplied by the normal ordered product.  As we mentioned above, the
normal ordered product induces an associative dot product in the
cohomology.  Therefore in order to prove that \(\ipco(z)\) is the
inverse to the picture changing operator in the cohomology all we have
to prove is that
\begin{equation}
  \label{eone}
  [\ipco]\cdot[\pco]=1
\end{equation}
where \([\phi]\) denotes the cohomology class of \(\phi\). Indeed, it
follows from the associativity and commutativity of the dot product
and \eq{eone} that
\begin{displaymath}
  [\ipco]\cdot\big([\pco]\cdot[\phi]\big)
 =[\pco]\cdot\big([\ipco]\cdot[\phi]\big)
       =\big([\ipco]\cdot[\pco]\big)\cdot[\phi]=[\phi].
\end{displaymath}
The identity (\ref{eone}) follows trivially from the fact that
\begin{equation}
  \label{noone}
  \no{\ipco\,\pco}=1.
\end{equation}
Note that although \eq{noone} is satisfied exactly without any BRST
trivial term, it does not mean that taking a normal ordered product
with \(\ipco\) is an inverse operation to taking a normal ordered
product with \(\pco\).  In order to prove this fact in cohomology we
had to use associativity but the normal ordered product is not
associative off shell.

\subsubsection{Strongly physical states}
Strongly physical states are closed states in BRST complex that have a
form of a primary matter state multiplied by a standard ghost
factor. In the Neveu-Schwarz sector these states can be represented by
a conformal field 
\begin{displaymath}
  \varphi=c\,\dl(\gm)\,V,
\end{displaymath}
where \(V(z)\) is a matter conformal field of dimension \(1/2\), or
equivalently by a state
\begin{displaymath}
  \ket\varphi=c_1\dl(\gm_\frac{1}{2})\,V_{-\frac{1}{2}}\ket{0}
\end{displaymath}
Taking the normal ordered product \(\no{\pco\,\varphi}\) is equivalent
to evaluating \(\pco_{g_{-\frac{1}{2}}}\ket{\varphi}\). 
\begin{equation}
  \label{pcono}
  \no{\pco\,\varphi}=c\,[G_{-\frac{1}{2}},V]+\gm\,V
\end{equation}
\begin{equation}
  \label{pcos}
  \pco_{g_{-\frac{1}{2}}}\ket\varphi=
           c_1\,G_{-\frac{1}{2}}V_{-\frac{1}{2}}\ket0
           +\gm_{\frac{1}{2}}\,V_{-\frac{1}{2}}\ket0
\end{equation}
One can easily see the analogy between these formulae and the
\(1|1\)-dimensional result
\begin{displaymath}
    \pco_D\,\d x\dl(\d\xi)\,f(x,\xi)
     =(\d x-\xi\,\d\xi)\,Df(x,\xi)+\d\xi\,f(x,\xi).
\end{displaymath}
We already mentioned the analogy between \(c\) and \(\d
x-\xi\,\d\xi\) in the previous section.  Here we continue this analogy
by identifying \(\gm\) with \(\d\xi\) and \(G_{-\frac{1}{2}}\) with
\(D\).  Again, the reasons for this analogy to take place will be
clarified in section~\ref{sec:conn}. 

The new matter vertex \([G_{-\frac{1}{2}},V]\), which appears in the
\eq{pcono} was originally presented as a result of the picture
changing operation on the vertex \(V\).  This interpretation was
particularly attractive since the state  \([G_{-\frac{1}{2}},V]\) has
conformal dimension one, which calls for an analogy with the bosonic
string. In the bosonic string strongly physical states have the form 
\begin{displaymath}
  \varphi_{\text{bose}}=c\,V_{\text{bose}},
\end{displaymath}
where \(V_{\text{bose}}\) is a dimension one primary matter
vertex. The corresponding state is annihilated by \(Q^{(0)}\), but not
\(Q^{(1)}\) or \(Q^{(2)}\) and therefore cannot be used as a
superstring state. Another problem with this state is that it is
annihilated by the inverse picture changing operator. 

It is rather peculiar that the two terms in \eq{pcono} or~(\ref{pcos})
play very distinct roles in the calculations.  The first term is the
only term to contribute to amplitudes with other strong physical
states but it vanishes under the inverse picture changing operation.
The second term does not contribute to the amplitudes but it is the
only one to reproduce the original state under the inverse picture
changing operation.

\section{From superstring to supermanifold}
\setcounter{equation}{0}
\label{sec:conn}

In this section we will tie together the results obtained so far by
showing the relation between the superstring BRST complex and the de
Rham complex of the super-moduli space.  For each \(n\)-string state
expressed as an element of an \(n\)-fold tensor product of the
superstring BRST complexes we will define a corresponding singular
form on the supermoduli space of decorated superconformal manifolds.
The even and odd dimensions of the resulting differential forms will
be determined by the gradings of the multi-string state.  We will show
that the action of the BRST operator on the multi-string state is
equivalent to taking the de Rham differential of the corresponding
form.  Similar result will be obtained for the picture changing
operators. 

\subsection{Superstring forms}
As it was shown in \cite{bel96a}, for any \(n\)-string state
\(\ket\Psi\) given as an element of the \(n\)-fold tensor power of the
BRST complex one can define a differential form on the moduli space of
decorated superconformal surfaces \(\P_{g,n}\).  Let \(\ket\Psi\) have
the superghost number \(-\deg(\bra\Sigma)+r|s\), then the corresponding
form will have degree \(r|s\) and its value on \(r\) even and \(s\)
odd vectors can be computed as
\begin{multline*}
  \Omega_{\Psi}(v_1,v_2,\dots,v_{r}|\hat{v}_1,\hat{v}_2,\dots,\hat{v}_{s})
   =\\ \bra{\Sigma}B(v_1)\,B(v_2)\cdots
   B(v_r)\,\dl(B(\hat{v}_1))\,\dl(B(\hat{v}_2))\cdots \dl(B(\hat{v}_s))\ket\Psi,
\end{multline*}
where \(\bra\Sigma\) is the state defined by the superconformal field
theory of ghosts and matter for the surface \(\Sigma\in\P_{g,n}\). The
superghost number of \(\bra\Sigma\) is determined by its genus and the
number of Neveu-Schwarz and Ramond punctures
\begin{displaymath}
  \deg(\bra\Sigma)=(3\,g-3+n_{\rm NS}+n_{\rm R})|
              (2\,g-2+n_{\rm NS}+\frac{1}{2}n_{\rm R}).
\end{displaymath}

Operators \(B(v)\) are the standard antighost insertions.  The
antighost incertions can be evaluated by representing a tangent vector
\(v\in T_\Sigma\P_{g,n}\) by a Schiffer variation
\(v=(v^{(1)},\cdots,v^{(n)})\), where
\(v^{(i)}=v^{(i)}_0(w_i)+\eta_i\,v^{(i)}_1(w_i)\) is a meromorphic
superconformal vector field in the \(i\)-th coordinate patch
\begin{displaymath}
  B(v)=\sum_{i=1}^n\Big(\oint \d w_i\; b^{(i)}(w_i)\,v^{(i)}_0(w_i)
                +\oint \d w_i\; \bt^{(i)}(w_i)\,v^{(i)}_1(w_i)Big).
\end{displaymath}
Similarly we define \(T(v)\) as
\begin{displaymath}
   T(v)=[Q,B(v)]=\sum_{i=1}^n\Big(\oint \d w_i\; L^{(i)}(w_i)\,v^{(i)}_0(w_i)
                +\oint \d w_i\; G^{(i)}(w_i)\,v^{(i)}_1(w_i)\Big),
\end{displaymath}
where \(Q=\sum_{i=1}^n Q^{(i)}\) is the BRST operator on the \(n\)-th
tensor power of the BRST complex.

It follows easily from the definition of the de Rham differential and
the conformal field theory property
\begin{displaymath}
  \dl_{v}\bra{\Sigma}=\bra{\Sigma}T(v),
\end{displaymath}
that the BRST operator acting on \(\Psi\) corresponds to taking the de
Rham differential of \(\Omega_\Psi\)
\begin{displaymath}
  \dd \Omega_{\Psi} = \Omega_{Q\,\Psi}.
\end{displaymath}
Similarly operator \(B(v)\) corresponds to the inner product and
\(T(v)\) to the Lie derivative
\begin{align}
  \inn_v\,\Omega_{\Psi} &= \Omega_{B(v)\,\Psi},\\
  \label{dlbdli}
  \dl(\inn_{\hat{v}})\,\Omega_{\Psi} &= \Omega_{\dl(B(\ov))\,\Psi},\\
  \label{lietv}
  \Lie_v\,\Omega_{\Psi} &= \Omega_{T(v)\,\Psi}.
\end{align}
The picture changing operator defined in section~\ref{sss:pco}, can be
generalized for the \(n\)-th tensor power of the BRST complex to
become
\begin{equation}
  \label{genpco}
  \pco_{\ov}=\frac{1}{2}\big(\dl(B(\ov))\,T(\ov)-T(\ov)\,\dl(B(\ov))\big)
            =\dl(B(\ov))\,T(\ov)+\dl(B(\ov))\,(T(\ov))^2.
\end{equation}
It follows from eqs.~(\ref{dlbdli}) and~(\ref{lietv})  that 
\begin{equation}
  \label{pnp}
  \pco_{\ov}\Omega_{\Psi}=\Omega_{\pco_{\ov}\Psi},
\end{equation}
where \(\pco_{\ov}\) in the right hand side should be evaluated
according to \eq{genpco}.

Finally a covector \(v^\vee\in T^*_\Sigma\P_{g,n}\) can be represented
by an \(n\)-tuple of meromorphic superconformal fields of dimension
\(3/2\),
\(v^{\vee{(i)}}=v^{\vee(i)}_0(w_i)+\eta_i\,v^{\vee(i)}_1(w_i)\), which
can be used to define an operator
\begin{displaymath}
  C(v^\vee)=\sum_{i=1}^n\Big(\oint \d w_i\; c^{(i)}(w_i)\,v^{\vee(i)}_0(w_i)
                +\oint \d w_i\; \gm^{(i)}(w_i)\,v^{(i)}_1(w_i)\Big),
\end{displaymath}
which corresponds to the exterior product operator 
\begin{align*}
  \ext_{\cv}\Omega_{\Psi}&=\Omega_{C(\cv)\,\Psi},\\
  \label{dlcdle}
  \dl(\ext_{\ocv})\Omega_{\Psi}&=\Omega_{\dl(C(\ocv))\,\Psi}.
\end{align*}

\subsection{States to fields correspondence}
States to fields correspondence is a familiar phenomenon for conformal
field theories.  Let us demonstrate how this correspondence works when
the states are taken from the BRST complex of the superstring.  In
this case it becomes more natural to associate an operator valued
differential form rather than an operator valued function with a state
which has a non-zero ghost number.

Let \(S_2\in\P_{0,2}\) be a super sphere with two NS punctures and
local coordinates given by \(w_1=z\), \(\eta_1=\th\) and \(w_2=-1/z\),
\(\eta_2=\th/z\), where \((z,\th)\) are the uniformizing coordinates
and the punctures are located at \((0,0)\) and \((\infty,0)\).  The
corresponding SCFT state defines a non-degenerate pairing on the BRST
complex. Using this pairing we can turn a state associated with a
3-punctured sphere with one inserted state \(\Phi\) into an operator
\(\Phi(S_3)\) with the matrix values 
\begin{displaymath}
  \bra{A}\Phi(S_3)\ket{B}=\braket{S_3}{A\otimes\Phi\otimes B},
\end{displaymath}
where \(\bra{A}\) is defined by 
\begin{displaymath}
  \braket{A}{B}=\braket{S_2}{A\otimes B}.
\end{displaymath}
Equivalently we can define an operator valued form 
\begin{displaymath}
  \bra{A}\Hat{\Omega}_\Phi\ket{B}=\Omega_{A\otimes\Phi\otimes B}.
\end{displaymath}
Let us define the map from the super complex plain \(\IC^{1|1}\) to
\(\P_{0,3}\) by mapping a point on \(\IC^{1|1}\) with coordinates
\((z,\th)\) to a sphere with punctures at \((0,0)\), \((z,\th)\)
and \((\infty,0)\) and the local coordinates given by 
\begin{align*}
  (w_1,\eta_1)&=(z',\th')\\
  (w_2,\eta_2)&=(z'-z-\th'\,\th,\th'-\th)\\
  (w_3,\eta_3)&=(-1/z',\th'/z')
\end{align*}
where \(z',\th'\) are the uniformizing coordinates.  Using this map we
can pull back the form \(\Hat{\Omega}_\Phi\) from \(\P_{0,3}\) to
\(\IC^{1|1}\) and define the form \(\Phi(\zz)\) associated with the
state \(\ket{\Phi}\).  

Let us now do some examples.
\begin{align*}
  \ket{\Phi}&=V(0)\ket0&\Phi(\zz)&=V(\zz)\\
  \ket{\Phi}&=c_1V(0)\ket0&\Phi(\zz)&=(\d z - \th\,\d\th)\,V(\zz)\\
  \ket{\Phi}&=c_1\dl(\gm_{\frac12})V(0)\ket0&\Phi(\zz)&=
  \d z\,\dl(\d\th)\,V(\zz),
\end{align*}
where \(V(z)\) is the matter only vertex and \(V(\zz)\) is defined as
\begin{displaymath}
  V(\zz)=V(z,\th)=V(z)+\th\,[G_{-\frac12},V(z)].
\end{displaymath}
This is exactly the same superconformal vertex operator as the one
used in FMS.  In particular \(V(\zz)\) satisfies
\begin{displaymath}
  [G_{-\frac12},V(\zz)]=DV(\zz)
\end{displaymath}
This construction explains the analogy between the BRST complex and
the de Rham complex of \(S^{1|1}\) which we observed several times in
the preceding discussion.

\section{Conclusion and outlook}

We have shown that the picture changing operation is not just a
bizzare feature of the superstring theory but is natural ingredient of
the geometrical integration theory on the supermanifolds.  This
interpretation allows one to rederive many properties of the picture
changing operators which were observed over the years in a consistent
manner.  The geometric description naturally explains the origin of
singularities that appear in the product of two picture changing
operators and allows to define the higher order picture changing
operators.  The higher order picture changing operators still present
some puzzles to be solved.  We introduced order \(n\) operators in
section~\ref{ss:pco} simply as symmetrized products of first order
picture changing operators associated with \(n\) linearly independent
odd vector fields \(\ov_i\)
\begin{displaymath}
  \pco_{\ov_1\cdots\ov_n}^{(n)}=\sum_{\sigma\in\mathfrak{S}_n}
  \pco_{\ov_{\sigma(1)}}\cdots\pco_{\ov_{\sigma(n)}}.
\end{displaymath}
There is a strong evidence that \(\pco_{\ov_1\cdots\ov_n}^{(n)}\) does
not change if the set of odd vector fields \(\{\ov_1\cdots\ov_n\}\) is
replaced by another set which spans the same \(n\)-dimensional space
at every point of the supermanifold.  It is relatively simple to
verify this hypothesis for \(n=1\) and \(n=2\) and for constant vector
fields with a straightforward calculation.  This suggests that the
higher order PCOs should be interpreted as an integration over the
subspace spanned by \(\{\ov_1\cdots\ov_n\}\).  This is easy to check
if the vector fields have vanishing Lie brackets with each other and
with themselves, but in general it is hard to define such ``partial
integration''.  The main problem is that by exponentiating odd vector
fields with non-vanishing Lie brackets one introduces a flow in even
directions as well as odd ones.

There still remain some puzzles with the off-shell generalization of
the picture changing operators.  It is not clear yet what is the
significance of the BRST exact terms in the product of two PCOs.  It
is likely that they may play a prominent role in the superstring field
theory.  

We did not analyze the relation between the picture changing operation
and the space-time supersymmetry.  In \cite{bel97} we demonstrated
that the SUSY algebra can be constructed geometrically without picture
changing, but the question whether the resulting algebra commutes with
the PCOs off-shell is still unclear.

Our analysis was done entirely within the holomorphic sector of the
superstring, but it should be straightforward to generalize it to the
full type II superstring.  Further understanding of the off-shell
structure of the type II superstring BRST complex will hopefully shed
some light to the long standing problem of constructing the full
string field action for the Ramond-Ramond sector.  Some progress in
this direction has been achieved recently in
\cite{Berk96,Berk97,Berk97a}.

\vspace{.6cm} \centerline{\Large\bf Acknowledgments} \vspace{.3cm} 

\noindent I would like to thank Louise Dolan, Paul Frampton, Walter 
Troost and Barton Zwiebach for their comments on the preliminary
version of this paper. Some calculations leading to the results
reported in this paper were verified using \texttt{OPEdefs}
\textsl{Mathematica}\footnote{\textsl{Mathematica} is a registered
  trademark of Wolfram Research.} package by Kris Thielemans.
\cite{OPEdefs}. 


\end{document}